\documentclass[showpacs,preprintnumbers,amsmath,amssymb]{revtex4}
\usepackage{epsfig}
\usepackage{graphicx}
\usepackage{dcolumn}
\usepackage{bm}
\newcommand{\be}{\begin{equation}}
\newcommand{\ee}{\end{equation}}
\newcommand{\ba}{\begin{eqnarray}}
\newcommand{\ea}{\end{eqnarray}}
\newcommand{\jj}{J/\psi}

\begin{document}
\preprint{CERN-PH-TH/2004-156; FNT/T-2004/14; ROME1-1386/2004}

\title{$J/\psi$ Absorption in  Heavy Ion Collisions II}
\author{L. Maiani}
\email{luciano.maiani@roma1.infn.it}
\affiliation{Universit\`{a} di Roma `La Sapienza', Roma, Italy; 
I.N.F.N., Sezione di Roma, Roma, Italy}
\author{F. Piccinini}
\email{fulvio.piccinini@pv.infn.it}
\affiliation{I.N.F.N. Sezione di Pavia and Dipartimento di 
Fisica Nucleare e Teorica, via A.~Bassi 6, I-27100 Pavia, Italy}
\author{A.D. Polosa}
\email{antonio.polosa@cern.ch}
\affiliation{Centro Studi e Ricerche ``E. Fermi'', via Panisperna 89/A-00184,
Roma, Italy}
\author{V. Riquer}
\email{veronica.riquer@cern.ch}
\affiliation{CERN, Department of Physics, Theory Division, Geneva, Switzerland}

\date{\today}

\begin{abstract} Using the methods introduced in a previous 
paper, we consider the dissociation of $J/\psi$ by the lowest-lying 
pseudoscalar and vector mesons in the hadronic fireball formed in heavy 
ion collisions, assumed to be a hadron gas at temperature $T$. 
Absorption by nuclear matter is accounted for as well. 
We compare with the S-U and Pb-Pb data presented by the 
NA50 Collaboration. From data 
at low centrality we find $T=165-185$~MeV, close to the 
predicted temperature of the transition to quark-gluon 
plasma and to the temperatures measured by hadron abundances. 
We extrapolate to higher centralities with the approximation 
of scaling the energy density of the fireball with the average 
baryon density per unit transverse area. Using the 
energy-temperature relation of the hadron gas made 
by the same pseudoscalar and vector mesons, the fall 
off of $\jj$ production shown by the NA50 data can 
be marginally reproduced only for the highest 
temperature, $T=185$~MeV. If we use the 
energy-temperature relation of a Hagedorn 
gas with limiting temperature $T_H=177$~MeV, 
predictions fall short from reproducing the data. 
These results suggest that a different mechanism 
is responsible for the $\jj$ suppression at 
large centralities, which could 
very well be the formation of quark-gluon plasma.


\end{abstract}
\maketitle

\section{Introduction}  In a recent paper~\cite{ioni} 
we have presented a theoretical study of the dissociation cross-section 
$\pi + \jj\to D^{(*)} +\bar D^{(*)}$ to analyze 
the absorption of $\jj$ in heavy ion collisions. 
The disappearance of the $\jj$'s created in the 
early stages of the collision was considered as mainly 
due to two subsequent processes:

${\bf 1.}$~The interactions of $\jj$ with the nuclear medium 
traversed during the interpenetration of the two heavy 
nuclei,with an absorption cross section per nucleon experimentally 
determined~\cite{nuclabs} from the inclusive $\jj$ 
production in $p+A$ collisions; 

${\bf 2.}$~The dissociation process: $\pi + \jj\to D^{(*)} +\bar D^{(*)}$ 
induced by the pions of the hadron gas formed in the heavy-ion collision, 
the so-called co-moving particles~\cite{Bjorken}.

In~\cite{ioni} the absorption curves thus obtained have been compared to a 
data analysis by the NA50 collaboration~\cite{na50}. Fitting the 
data at low centrality we found a value 
for the inner temperature of the pion gas of about $T=225\pm 15$~MeV. 
Extrapolation of the absorption curve at large centrality was done 
by assuming an increase of the energy density deposited in the fireball 
proportional to the average number of nucleons per unit of overlapping 
area~\cite{Bjorken}, a quantity which increases with decreasing impact 
parameter. The results did not seem to reproduce the fast increase of 
absorption observed by NA50, thus lending some support to the idea 
that in large centrality collisions a quark-gluon plasma phase is 
produced, which would impair the formation of the $\jj$ because of 
QCD Debye screening~\cite{screening}.As already noted in~\cite{ioni}, 
however, for this method of analysis 
to be accurate one should go beyond the approximation of a simple 
pion gas and consider the more realistic case in which further 
particles/resonances appear in the fireball at thermal equilibrium, 
which can appreciably contribute to $\jj$ dissociation. 

In the present paper, we extend our previous work in 
several directions. First,  we include the $\jj$ dissociation 
cross sections by the lowest lying pseudoscalar 
and vector mesons (Sect.~II). We use the Constituent 
Quark Model~(CQM)~\cite{cqm} as before, with couplings 
computed with flavour SU(3) symmetry, and nonet symmetry for the 
vector mesons. Real particle masses are used for reaction thresholds 
and to compute particle abundances in the hadron gas. 

After defining the different lengths that characterize 
the collision (Sect.~III), the absorption lengths vs. temperature 
are computed in the heath bath made by the hadron gas of 
pseudoscalar and vector mesons at a given 
temperature $T$ (Sect.~IV). Not unexpectedly~\cite{muller}, the largest 
correction to the pion gas situation is due to the vector mesons. 
Notwithstanding the considerably larger mass, they contribute to 
$\jj$ absorption about as much as pions, for two reasons: (i)~dissociation 
reactions have very low or no threshold, thus making all particles useful 
as opposed to pions, which are effective only above a relatively 
high-energy threshold; (ii)~the large spin and flavour multiplicities 
(a factor of 24 with respect to 3 for pions, for the complex 
$\rho + \omega + K^{*} + \bar{K}^*$). 

The total inverse absorption length, 
$\Sigma_{i} \langle\rho_{i}\sigma_{i}\rangle_{T}$, 
is considerably increased with respect to the pure pion gas. 
Correspondingly, we find a substantial decrease in the estimated temperature 
of the fireball in low centrality collisions, which turns out to be now 
in the range $165-185$~MeV (see Fig.~\ref{f:abscombi}). Extrapolation 
to higher centrality is done using the relation between energy density 
and temperature appropriate to the hadron gas we are considering and it 
leads to increasing temperatures with increasing centrality up to 
$180-200$~MeV at zero impact parameter, $l=2R=13$~fm for 
Pb~($l$ is the linear size of the fireball and $R$ the nuclear radius). 
The corresponding hardening of the attenuation curves, however, do not 
really reproduce the falling of $\jj$ production in this region, as 
shown by Fig.~\ref{f:abscombi}, supporting the idea of a change of 
regime, this time in a more realistic region of the thermodinamical 
parameters with respect to~\cite{ioni}. 

In the range of temperatures we have found, one expects to deal with 
a hadron gas of increasing complexity approaching the 
Hagedorn gas~\cite{hagedorn}, with an infinite number of resonances 
and a level density exponentially increasing with mass. This situation 
is known to give rise to a limiting temperature~\cite{rafelski}, 
which was interpreted in~\cite{Cab&Par} as the temperature at which 
the transition to a quark-gluon plasma phase starts to take place. 
In Sect.~V, starting from the fit at low centrality, we extrapolate 
to higher centrality with the energy-temperature relation appropriate 
to the Hagedorn gas, with a Hagedorn temperature $T_{H}=177$~MeV. 
This temperature is consistent with the transition temperature estimated 
in QCD lattice calculations~\cite{lattice}, with the empirical 
density of hadron levels up to $2$~GeV, as estimated by~\cite{rafelski}, 
and with the temperatures determined at SPS and RHIC from particles 
relative abundances at freeze-out~\cite{antinori}. We find that the 
increase of the energy deposited in the collision does not lead 
anymore to an increase of temperature, hence of the opacity: the 
extra energy deposited in the fireball for increasing centrality 
goes into an increase of the thermodinamical degrees of freedom, 
which ultimately should lead to the phase transition. 

The curve corresponding to the absorption by the Hagedorn gas  
falls definitely short to explain the further decrease of $\jj$ 
abundance, as observed by NA50. It is tempting to interpret the 
discrepancy shown in Fig.~\ref{f:hagedorn}  as a signal of 
quark-gluon plasma formation. 
However, caution is required, due to the unavoidable truncation 
we have made in the calculation of the opacities. While higher 
resonances are expected to contribute less and 
less, their cumulative effect could resum to a large contribution to the 
dissociation cross-section and lead to an increase in opacity even for 
the very small temperature increases allowed by the vicinity of the 
limiting temperature.

Sect.~VI contains conclusions and outlook.We find it very 
satisfactory that a microscopic calculation based on the CQM 
and the observed opacities produce values of the temperature 
that are consistent with the measured freeze-out temperature 
in ion collisions and are just at the border of the transition 
temperature predicted by QCD lattice calculation and by the 
Hagedorn hadron gas. A point borne out clearly by our analysis 
is that the picture would be 
considerably more clear if the relative normalization of S-U vs. Pb-Pb 
data was better understood~\cite{na50web}. In this respect, further 
measurements at low energy, aimed at resolving the experimental issue of 
the relative normalization, could be very useful. 

All in all, the results shown in Figs.~\ref{f:abscombi}~(a),(b) 
or, even more, in Fig.~\ref{f:hagedorn} are quite impressive. 
They seem to us very suggestive that a new mechanism to suppress 
$\jj$ is setting in at large centralities. This could very well 
be the formation of quark gluon plasma. 
It would be extremely important to correlate other 
signals to the present one, in order to get to a definite 
conclusion about the presence of a phase transition.

In this paper we correct few mistakes which affected our 
analysis in~\cite{ioni}. The pion cross sections presented here 
supersede those in ~\cite{ioni}, where we found a trivial numerical 
mistake; the average lenght traversed by the $\jj$ in the fireball 
was incorrectly estimated to be $(6/10){\it l}$ rather than $(3/8){\it l}$. 
All together these corrections would increase the temperature given 
in~\cite{ioni} by $15-20$~MeV, leaving all conclusions unchanged. 
Finally, a misprint led to the factor $2\pi/3$ in Eq.~(13) of 
Ref.~\cite{ioni}, to be replaced by $\pi/2$.

\section{Cross section computations}

The evaluation of the cross sections for the processes
$(\pi,\eta,K,\rho,\omega,K^{*},\phi)+\jj\to D^{(*)}_{(s)}D^{(*)}_{(s)}$
proceeds through the computation of diagrams very similar to 
those listed in~\cite{ioni}. Such tree level diagrams 
involve effective tri-linear 
$g_3=(\pi,\eta,K,\rho,...)D^{(*)}_{(s)}D^{(*)}_{(s)}$ or  
$g_3=\jj D^{(*)}_{(s)}D^{(*)}_{(s)}$
and four-linear
$g_4=(\pi,\eta,K,\rho,...)\jj D^{(*)}_{(s)}D^{(*)}_{(s)}$
couplings which we can estimate using the CQM model,
originally devised to compute exclusive
heavy-light meson decays and tested on a quite large number of
such processes~\cite{cqm}.

\begin{figure}[ht]
\begin{center}
\epsfig{
height=4.0truecm, width=10.truecm,
        figure=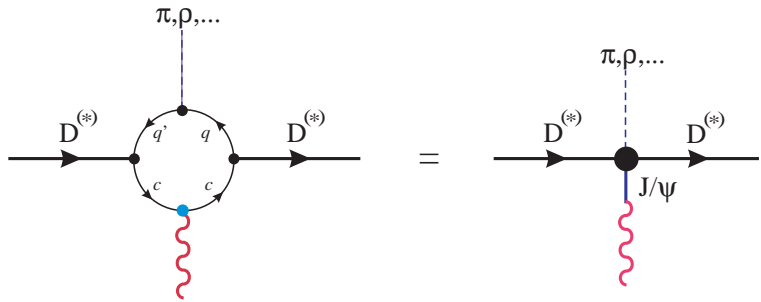
}
\caption{\label{f:loop} \footnotesize 
Basic diagrammatic equation to compute $g_3$ and $g_4$ couplings.
}
\end{center}
\end{figure}

CQM is based on an effective Lagrangian which incorporates the 
heavy quark spin-flavor symmetries and the chiral symmetry in the
light sector. In particular, it contains effective vertices
between a heavy meson and its constituent quarks~(see
the vertices in l.h.s. of Fig.~\ref{f:loop}) whose 
emergence has been shown to occur when applying 
bosonization techniques to Nambu-Jona-Lasinio interaction 
terms of heavy and light quark fields~\cite{ebert}.
On this basis we believe that CQM is a more solid approach 
to the computation of $g_3,g_4$ if compared to the various
methods available in the literature, often based on SU(4) symmetry
(for a review see e.g.~\cite{barnes}).

In Fig.~\ref{f:loop} we show the typical equation which has to be solved 
in order to obtain $g_4(g_3)$ in the various cases at hand: on the r.h.s.
we represent the effective four-linear couplings to be
used in the cross section calculation (to obtain the tri-linear
coupling we suppress either the $\jj$ line or the dashed 
line of the light particles); the effective interaction at 
the meson level (r.h.s.) is modeled as an interaction at the 
quark-meson level (l.h.s. of Fig.~\ref{f:loop}).

The $\jj$ is introduced using a Vector Meson Dominance~(VMD) Ansatz:
in the effective loop on the l.h.s. of Fig.~\ref{f:loop} we have a vector 
current insertion on the heavy quark line~$c$ while on the r.h.s. the
$\jj$ is assumed to dominate the tower of $J^{PC}=1^{- -}$, $c\bar c$ states
mixing with the vector current (for more details see~\cite{pioni}). 
Similarly, vector particles coupled to the light quark component
of the heavy mesons $\rho,\omega$, when $q=(u,d)$, or $K^*,\phi$,
when one or both light quarks 
are $q=s$, are also taken into account using VMD arguments. 
The pion and other pseudoscalar fields
have a derivative coupling to the light quarks of the Georgi-Manohar 
kind~\cite{manogeo}.

Once established the form of the effective vertices occurring in 
the loop diagram on the l.h.s. 
of Fig.~\ref{f:loop}, one has just to compute it. 
The momenta running in the loop are limited
by two Schwinger cut-off's: one in the ultraviolet and one 
in the infrared.
The ultraviolet cut-off $\Lambda$ has been set to the chiral expansion scale 
$\Lambda_\chi\simeq 4\pi f_\pi$ while the infrared $\mu$ prevents loop
momenta to 
access the confinement energy region (CQM does not include any 
confining potential).
To give a flavor of the kind of calculations involved, consider that
the loop integral for the four-linear coupling $\rho\jj D^{(*)}D^{(*)}$ is written as follows:
\begin{eqnarray}
\label{eq:loop}
&&(-1)\sqrt{Z_H m_H Z_{H^\prime} m_{H^\prime}}
\times N_c\int\frac{d^4l}{(2\pi)^4 i}\times\nonumber\\
&& {\rm Tr}\left[(-i\bar
H^\prime(v'))\frac{1}{v'\cdot l+\Delta}(i\frac{m^2_J}{f_J}\eta\!\!\!/)
\frac{1}{v\cdot l+\Delta}(-iH(v))\frac{1}{l\!\!/
-m}(i\frac{m_\rho^2}{f_\rho}\epsilon\!\!/)\frac{1}{l\!\!/+q\!\!\!/
-m}\right],
\end{eqnarray}
where $H$ and $\bar H^\prime$ represent the heavy-light external meson fields
labeled by their four-velocities $v,v^\prime$; the Feynman rules for 
their couplings to constituent quarks, in this case the 
$\sqrt{Z_H m_H Z_{H^\prime} m_{H^\prime}}$ coupling factor, 
are discussed in~\cite{cqm}. 
The heavy quark propagators
(those obtained by the standard Dirac propagator in the limit 
of very heavy $Q=b,c$ mass) 
are also labeled by $v,v^\prime$ while we have the usual expression 
for the light propagators. The parameter $\Delta$ appearing in the heavy 
propagator is defined to be $\Delta=M_H-m_Q$, the mass of the heavy-light 
meson minus the mass of the heavy quark contained in it. $\Delta$ is the
the main free parameter of the model. 
It varies in the range 
$\Delta=0.3-0.5$~GeV for $u,d$ light quarks and $0.5-0.7$~GeV 
for strange quarks~\cite{eff0}. Varying $\Delta$ allows to estimate 
the theoretical error.

The $\rho$ is coupled via its VMD coupling
to the light quarks, $\epsilon$ being its polarization. The $\jj$, with 
polarization $\eta$, is also coupled via VMD to the heavy quarks 
($\eta$ appears in the trace between two 
heavy quark propagators, $\epsilon$ appears between two 
light quark propagators). 
In front of this expression we have the fermion loop factor.

The way to regularize and compute such integrals is discussed 
in~\cite{lapol}. The trace computation in~(\ref{eq:loop})
will introduce a number of scalar combinations of the momenta 
and polarizations of the external particles. Each of these combinations 
will be weighted by a scalar integral which amounts to a numerical 
factor: what we call the coupling.
Actually such scalar integrals depend on the energy of the
$\rho$. In general the expressions
obtained in the $\sigma_{\rho\jj}$ computation 
appear to be quite complicated functions of $E_\rho$
if compared to those 
obtained when studying only $\jj$ interactions 
with pions~\cite{pioni}. 
On the other hand, having in mind a hadron gas
at a temperature $T\approx 170$~MeV, we restrict our study 
to the low energy interval $E_\rho\simeq 770-1000$~MeV and
rely on the fact that the Boltzmann factor exp$(-E/T)$ will cut 
off the high energy tails and work as a high energy form factor 
in the thermal averages $\langle \rho \sigma\rangle_T$ (we have
explicitly tested that this occurs).

The couplings $\rho D^{(*)}D^{(*)}$, are computed in the same 
framework by simply dropping the $\jj$ line in the
diagrammatic equation of Fig.~\ref{f:loop}, as mentioned above.

The $\omega$ particle contribution 
is introduced in the thermal average by simply multiplying 
the $\rho$ cross-section by a factor of $4/3$ (nonet symmetry). 
The amplitudes for $K^*$ are deduced from those of the $\rho$ by 
SU(3) symmetry. However, computing the $K^*$ cross-sections, 
where a $D_s$ meson will be produced in the final state, 
we take into account exact masses and thresholds. 
Similarly $\eta$ and $K$ contributions are computed by SU(3) 
symmetry from the pion couplings~\cite{pioni} but accounting 
for different kinematical thresholds.  The calculation of the 
$\phi$ contribution is 
performed by replacing the constituent light quark mass, 
$m=300$~MeV for $q=(u,d)$, with a constituent strange quark mass $m=500$~MeV.
Such an operation requires accordingly the modification of the 
$Z_H$ couplings (see Eq.~(\ref{eq:loop})) and of the 
infrared cutoff, as described in~\cite{eff0}.

\begin{figure}[ht]
\begin{center}
\epsfig{
height=6.truecm, width=12.truecm,
        figure=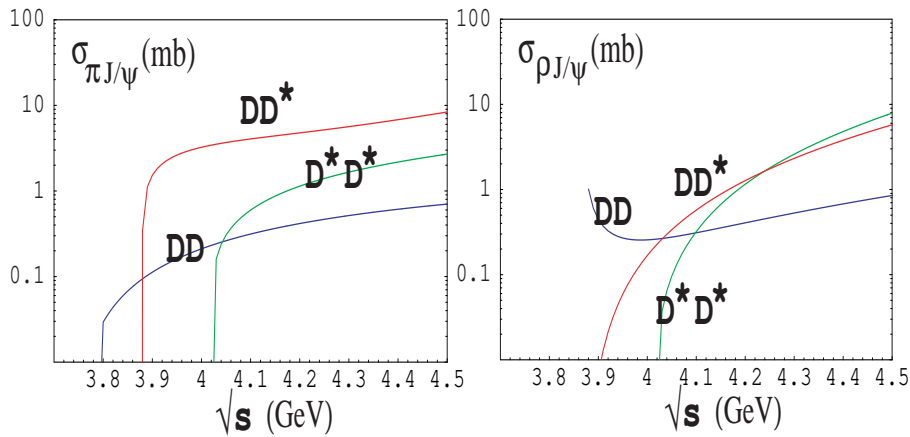
}
\caption{\label{f:xsects} \footnotesize 
The cross sections for the processes
$(\pi^+,\rho^+)+J/\psi\to D^{(*)}\bar D^{(*)}$ versus energy.
}
\end{center}
\end{figure}

The final results for $\sigma_{\pi\jj}$ and $\sigma_{\rho\jj}$ 
are displayed in Fig.~\ref{f:xsects}. We give the cross sections 
for $\pi^+$ and $\rho^+$.

\section{The geometry of the collisions}

\begin{figure}[ht]
\begin{center}
\epsfig{
height=6.truecm, width=12.truecm,
        figure=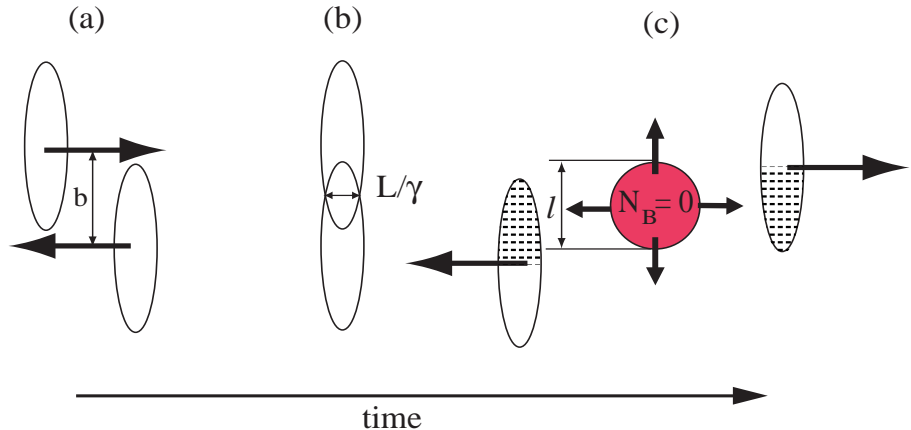
}
\caption{\label{f:geo} \footnotesize 
Time evolution of the collision of two heavy nuclei seen in the 
c.o.m. frame.
A fireball with baryon number $N_B=0$
is left behind the two receding nuclei. The fireball is expected 
to thermalize at the temperature $T$.
The nucleons belonging to the nuclear overlapping regions 
fly apart as unbounded  particles.
The dimensions $L/ \gamma$ and $l$ are respectively the linear dimension of
the nuclear column the $\jj$ has to traverse during the interpenetration 
of the two nuclei and the linear size of the fireball.
}
\end{center}
\end{figure}

The geometry of the heavy ion collision is shown schematically 
in Fig.~\ref{f:geo}, 
which depicts the time-evolution in the center of mass frame. 

The impact parameter, $b$, is defined, as usual, as the transverse distance 
of the centers of the two nuclei. We consider the $\jj$ to be 
created with Feynman's $x \approx 0$, during the overlap 
of the two nuclei. These particles have to overcome absorption 
from the column density of nucleons. In the center of mass frame 
the length of the column is $L/\gamma$. In the same frame, the 
density of nucleons is $\rho_{\rm nucl.} \cdot \gamma$,
so that the absorption factor is Lorentz invariant and given 
by ${\rm exp}(-\rho_{\rm nucl.}\sigma_{\rm nucl.}L)$. 
This is the same formula used in Ref.~\cite{na50}.  
$\sigma_{\rm nucl.}$, the nuclear absorption cross-section, 
has been determined by NA50 from the behaviour of the cross-section 
of $p + A \to \jj$~+~anything as function of A. 
We take from Ref.~\cite{nuclabs}:
\begin{eqnarray}
\label{eq:sigmanucl}
&&\sigma_{\rm nucl.}= 4.3 \pm 0.3~{\rm mb}\\
&&\rho_{\rm nucl.}= 0.17~{\rm fm}^{-3}.
\end{eqnarray}

After collision, nuclear matter appears in part as a 
cloud of free (wounded) nucleons from the initially overlapping parts 
of the nuclei, represented as dotted regions in Fig.~\ref{f:geo}~(c), 
in part as forward going fragments from the non overlapping regions. 
The NA50 Collaboration has installed a Zero-Degree-Calorimeter (ZDC) 
which measures the energy of the forward going fragments, proportional 
to their nucleon number and therefore to their size, thereby determining 
the value of the impact parameter, $b$, for each collision~\cite{na50}. 
The relation between $L$ and $b$ has been given in~\cite{kharzeev} 
in the framework of the Glauber theory.

In Fig.~\ref{f:geo} we show the hadron fireball 
produced by the central collisions of the 
interacting nucleons~\cite{Bjorken} (the {\it comoving particles}). 
The fireball has a transverse dimension, $l$, approximately 
equal to the length of the overlapping region:
\begin{equation}
\label{eq:elle}
{\it l}=2R-b,
\end{equation}
where $R$ is the nuclear radius. 
If we take, for simplicity, a spherical fireball, the average 
length that a $\jj$ has to traverse before leaving it is
$(3/8){\it l}$, so that the attenuation factor due to absorption 
by the comoving particles is:

\begin{equation}
\label{eq:abscomov}
{\it A}_{\rm comoving}\propto {\rm exp}\left[-\Sigma_i \langle
\rho_i\sigma_i\rangle \frac{3}{8}l\right],
\end{equation}

\noindent the subscript $i$ labels the species of hadrons making up the fireball, 
${\rho}_i$ the number density of the effective (i.e. above threshold)
particles and ${\sigma}_{i}$ the corresponding $\jj$ dissociation 
cross-section. Brackets indicate an average over the energy distribution 
in the fireball. As noted before, we can express the nuclear absorption 
length, $L$, as a function of $b$~\cite{kharzeev} and therefore, 
using~(\ref{eq:elle}), as a function of ${\it l}$. Putting all 
together, we write the attenuation of the $\jj$ as a function 
of $\it l$ according to:

\begin{equation}
\label{eq:attot}
A=N\times {\rm exp}[-\rho_{\rm nucl.}\sigma_{\rm nucl.} L(l)]\times
{\rm exp}\left[-\Sigma_i \langle \rho_i \sigma_i\rangle\frac{3}{8}l
\right],
\end{equation}

\noindent where $N$ is an appropriate normalization constant.
As discussed in Ref.~\cite{Bjorken}, the energy density deposited 
in the fireball is proportional to the average flux of nucleons 
participating to the collision, i.e., the number of nucleons per 
unit transverse area. This quantity increases with decreasing 
impact parameter. A simple estimate of the ratio of the energy 
density for two different values of $b$ was given in ~\cite{ioni} 
according to:

\begin{equation}
\label{eq:geom}
\frac {\epsilon (b_2/R)}{\epsilon(b_1/R)}=\frac{g(b_2/R)}{g(b_1/R)},
\end{equation}

with the geometrical factor $g(b/R)$ given 
by~\cite{ioni}:

\begin{equation}
\label{eq:gfact}
g(b/R)=\frac{\pi}{2}\frac{(1-b/2R)^2 (1+b/4R)}
{{\rm arccos}(b/2R)- (b/2R)\sqrt{1-b^2/4 R^2}}.
\end{equation}

\section{Thermal averages and absorption in the resonance gas}

The fireball depicted in Fig.~\ref{f:geo}~(c) is expected to quickly 
thermalize (see \cite{Bjorken}), giving rise, at least for low 
centrality collisions, to a hadron gas at temperature $T$. 
The thermalisation hypothesis is supported by the data from 
SPS and RHIC which show abundances and momenta distributions 
at freeze-out compatible with a temperature $T=170-180$~MeV, 
see e.g. Ref.~\cite{rafelski} and Ref.~\cite{antinori} for a 
review of recent data.

In Ref.~\cite{ioni} we have assumed the thermalised fireball 
to be a pion gas. We consider now a more general case of a 
boson gas made by the lowest-lying pseudoscalar and vector 
mesons, retaining, for simplicity, the assumption of vanishing 
chemical potential. What matters, for our considerations, are 
particle densities above the dissociation threshold. Assuming 
the $\jj$ to be at rest, this means particles with energy larger 
than $E_{\rm th.}$:

\begin{equation}
\label{eq:Ethr}
E_{\rm th.}=\frac{(M_{fin})^2-(M_{\jj})^2-m^2}{ 2 M_{\jj}};
\end{equation}

\noindent $m$ is the projectile particle mass and $M_{fin}$ the sum of the 
final particles masses. Explicitly, we have:
\begin{equation}
\label{eq:rho}
\rho(T)=\frac{N}{2\pi^2}\int_{E_{\rm th.}}^{\infty} dE \frac{pE}{e^{E/kT}-1}.
\end{equation}

\begin{figure}[ht]
\begin{center}
\epsfig{
height=7.truecm, width=8.truecm,
        figure=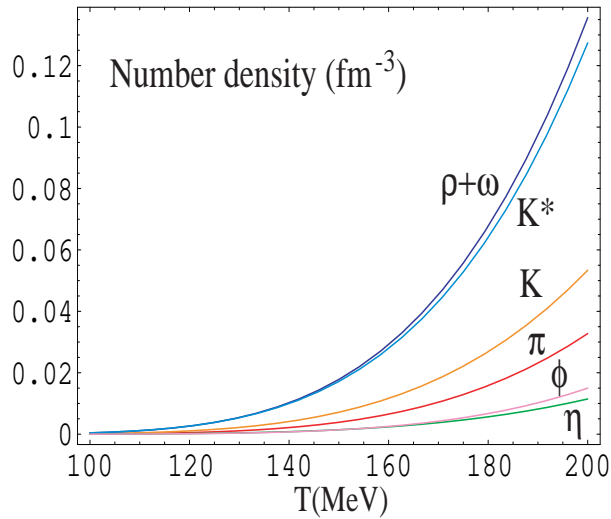
}
\caption{\label{f:numden} \footnotesize 
Number densities of the hadron gas particles $h$
being above the kinematical threshold 
for the  $h\jj\to D_{(s)}^{(*)}\bar D_{(s)}^{(*)}$ reaction.
}
\end{center}
\end{figure}

\noindent $N$ is the total multiplicity 
(spin times charge, $N=3,9$ for pion and for $\rho$, 
respectively) and $p=\sqrt{E^2-m^2}$. 
We report in Fig.~\ref{f:numden} and in Tables I and II the number 
densities above threshold for the particles considered. 
In Table I, we give in parenthesis the total number density 
of pions as well.

The energy density associated to the various particles 
are listed in Tables III to V. In this case the sum over 
the full energy range is understood:
 
\begin{equation}
\label{eq:enerd}
\epsilon(T)=\frac{N}{2\pi^2}\int_{m}^{\infty} dE \frac{pE^2}{e^{E/kT}-1}.
\end{equation}

The contribution of the other particles is not at all negligible 
with respect to pions, even at temperatures as low 
as $T=150$~MeV,~\cite{rafelski}, particularly in terms of 
the energy density. 

\begin{figure}[ht]
\begin{center}
\epsfig{
height=6.truecm, width=12.truecm,
        figure=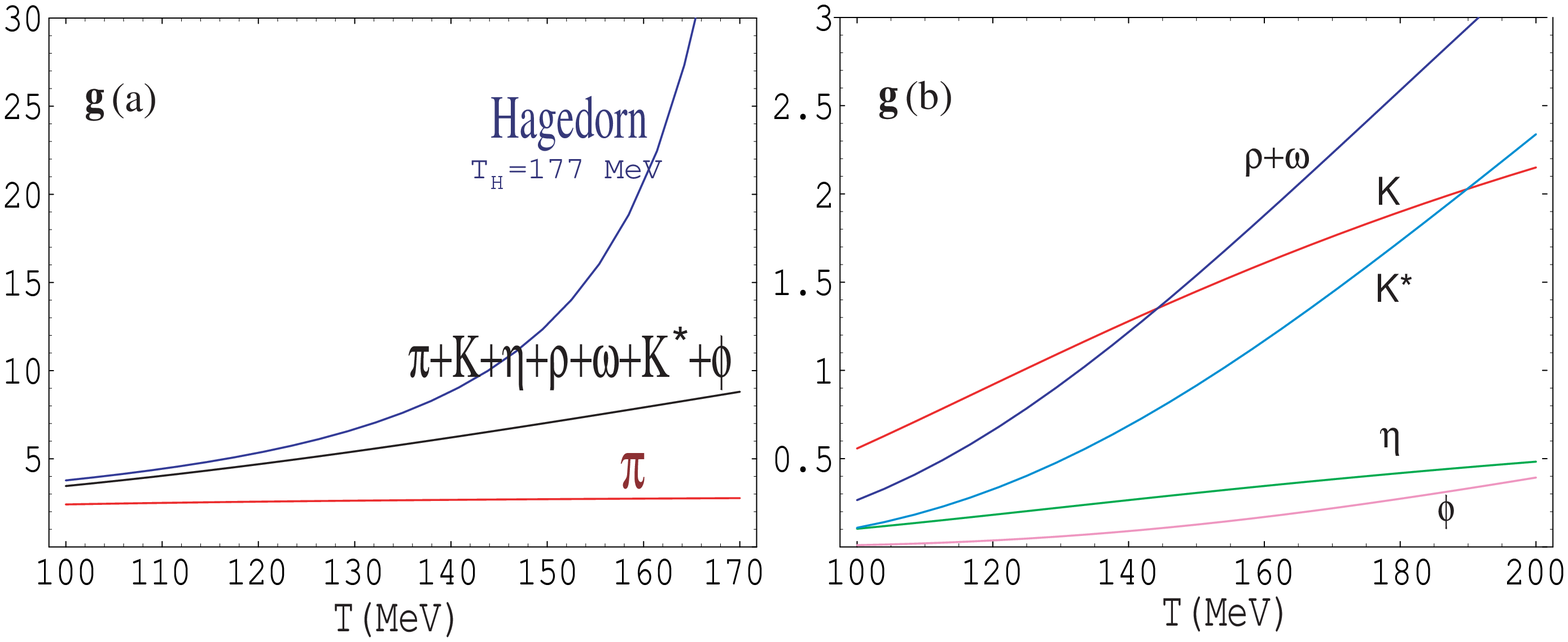
}
\caption{\label{f:gicombi} \footnotesize 
The effective number of degrees of freedom $g$ active at a particular
temperature. In panel (a) we show the dramatic increase of $g$ for 
the Hagedorn gas with limiting temperature $T_H=177$~MeV, 
see~(\ref{eq:epsilonHag}). 
In panel~(b) we specify the contribution of 
the different components of the hadron gas.
}
\end{center}
\end{figure}

We report in Figs.~\ref{f:gicombi}~(a,b) 
the energy densities divided by
$\epsilon_0=T^4\pi^2/30$. The ratio, $g$, gives the effective number 
of degrees of freedom which are active at the particular temperature. 
Pions give $g=3$ already at $T=150$~MeV (Fig.~\ref{f:gicombi}~(a)), 
which is increased to 
$g=7-10$ by the other particles, for $T=150-180$~MeV. 
In Fig.~\ref{f:gicombi}~(b) 
we specify the contributions of the different particles, 
with Kaons ($N=4$) and the lowest lying vector mesons, $\rho$ and 
$\omega$ ($N=12$) providing the dominant contributions.

Thermal averages of the product $\rho\cdot \sigma $ 
give the inverse of the absorption length due to each particle species, $x$, 
and are done with analogous formulae:

\begin{equation}
\label{eq:rhosigma}
\langle \rho \cdot \sigma_{x + \jj \to D^{(*)}D^{(*)}}\rangle_T=
\frac{N}{2\pi^2}\int_{E_{{\rm th.}}}^{\infty} dE
\frac{pE\sigma(E)}{e^{E/kT}-1}.
\end{equation}

\begin{figure}[ht]
\begin{center}
\epsfig{
height=7.truecm, width=8.truecm,
        figure=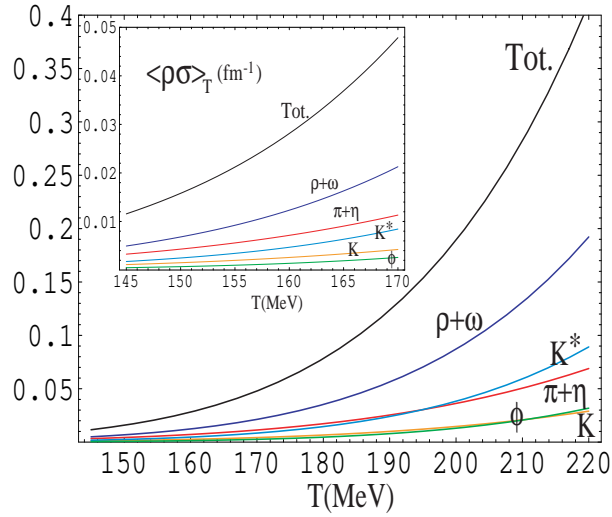
}
\caption{\label{f:absleng} \footnotesize 
The inverse absorption lengths as a function of temperature.
}
\end{center}

\end{figure}

We give the values of the inverse absorption lengths as function of 
temperatures in Table V and Fig.~\ref{f:absleng}. 
$\rho$ and  $\omega$ give larger contributions than the pions. 
This is due to the absence of threshold in the dissociation 
reaction, which makes particles of all energies to be effective, 
and to the large multiplicity.

\begin{figure}[ht]
\begin{center}
\epsfig{
height=7.truecm, width=15.truecm,
        figure=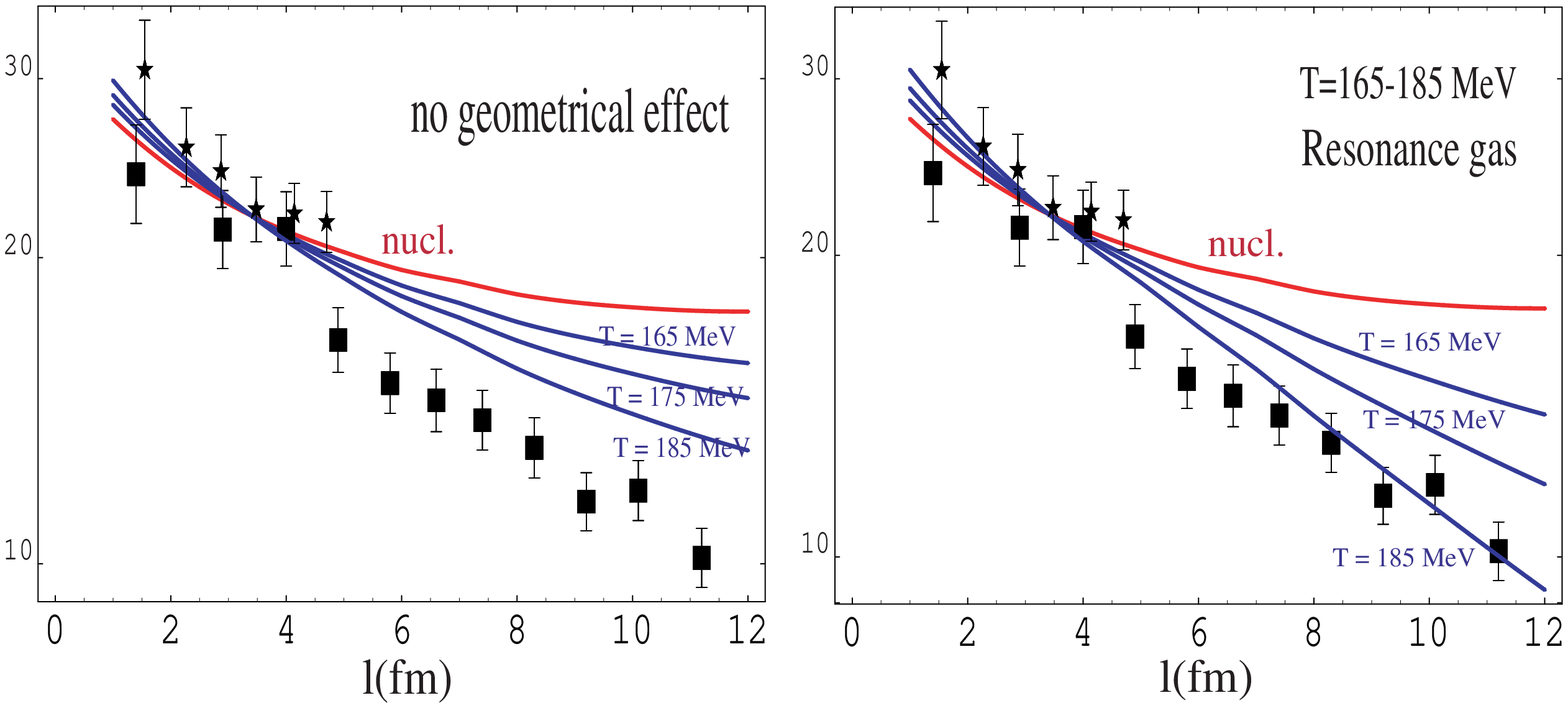
}
\caption{\label{f:abscombi} \footnotesize 
The absorption curves are compared to the most recent data analysis of
the $\jj$ yield in S-U~(star) and Pb-Pb~(box) collisions~\cite{na50web}. 
Left panel: the absorption by comoving particles is described by a 
single temperature; the effect of nuclear absorption alone is shown 
by the upper curve (since the relation between $L$ and ${\it l}$ 
is non-linear, the nuclear absorption does not give rise to a simple 
exponential, unlike the isothermal absorption by comoving particles).
Right panel: the temperature increase 
due to the geometrical effect described in Sect.~III is 
accounted for with the energy-temperature relation of the hadron gas. 
The temperatures indicated refer to ${\it l}=3.4$~fm. 
}
\end{center}

\end{figure}

The absorption lengths thus computed are inserted in the absorption 
master formula~(\ref{eq:attot}). We compare 
in Fig.~\ref{f:abscombi}, 
the absorption curves thus obtained with the 
NA50 data in S-U and Pb-Pb collisions, 
for $T=165-185$~MeV~\cite{nota2d}. 
The normalization constant in~(\ref{eq:attot}) 
is chosen to fit the data around $l=3$~fm. The data for $l\leq 4$~fm favour 
temperatures of this order.
In the same figure, we show the curve corresponding to the pure 
nuclear absorption. 

To extrapolate to higher centralities, we keep 
into account the increase in the energy density deposited 
in the collision as explained in Sect.~III. 
The  increase in temperature for increasing centrality makes 
the absorption to increase, as shown in Fig.~\ref{f:abscombi}~(b),
with respect to Fig.~\ref{f:abscombi}~(a) where the geometrical effect is
neglected.
In Fig.~\ref{f:abscombi}~(b) 
we label the curves with the value of the temperature 
at $l=3.4$~fm; the temperature increases along the 
curves, up to $T\approx 180-200$~MeV for $l=11$~fm.

The inclusion of the lowest lying resonances improves greatly the consistency 
of the whole picture, bringing the estimated 
temperatures below the temperature expected for the phase 
transition and closer to the temperatures measured at freeze-out, 
from the abundances of the different hadrons. The curve labeled with $T=175$~MeV, 
which has $T$ varying from $175$ to about $190$~MeV, fits well the data for 
low centrality but still falls short from reproducing the observed drop in 
the production of the $\jj$ above, say, ${\it l}=5$~fm. The curve with initial temperature
$T=185$~MeV fits the low centrality data as well and follows closer 
the data at large centrality. However the temperature exceeds $200$~MeV 
at $l\simeq 11$~fm which is likely too a high value for a hadron gas (see next Section).

The increase in temperature due to the increase in energy density 
that we find is definitely less pronounced than what found for 
the pion gas in \cite{ioni}. The reason is that in the resonance 
gas the number of degrees 
of freedom increases appreciably with temperature. The extra 
energy density provided has to be shared among more and more 
degrees of freedom and the temperature increases less than 
with a fixed $\epsilon= CT^4$ power law. This behaviour 
begins to reproduce what is expected in the case of a 
Hagedorn gas, with an exponentially increasing density 
of resonances per unit interval of 
mass~\cite{hagedorn},\cite{rafelski},\cite{Cab&Par}, which we turn now to consider.
\begin{table}[htdp]
\begin{center}
\begin{tabular}{|c|c|c|c|c|c|}
\hline
$T$~(MeV) & $\rho^\pi~(\rho^\pi_{Tot})$~(fm$^{-3}$) & $\rho^\eta$~(fm$^{-3}$) 
& $\rho^K$~(fm$^{-3}$) & $\langle \rho\sigma\rangle_T^{\pi+\eta}$~(fm$^{-1}$) &
$\langle \rho\sigma\rangle_T^{K}$~(fm$^{-1}$) \\
\hline
$150$&$0.004~(0.12)$&$0.001$&$0.006$&$0.004$&$0.001$\\
$165$&$0.008~(0.17)$&$0.003$&$0.012$&$0.009$&$0.003$\\
$180$&$0.016~(0.22)$&$0.006$&$0.022$&$0.017$&$0.007$\\
$195$&$0.028~(0.29)$&$0.010$&$0.039$&$0.030$&$0.012$\\
$210$&$0.045~(0.37)$&$0.016$&$0.062$&$0.050$&$0.021$\\
$225$&$0.069~(0.47)$&$0.024$&$0.095$&$0.080$&$0.034$\\
$240$&$0.102~(0.57)$&$0.035$&$0.139$&$0.121$&$0.052$\\
\hline
\hline
\end{tabular}
\caption{
Number densities and inverse absorption lengths for the pseudoscalar
particles in the gas. The number densities are relative to those 
particles which are over the kinematical threshold to open the 
$D^{(*)}_{(s)}\bar D^{(*)}_{(s)}$ channel. 
In the case of pions, we 
report in parenthesis their total number density (i.e., the one
including pions below threshold).
} 
\end{center}
\end{table}
\begin{table}[htdp]
\begin{center}
\begin{tabular}{|c|c|c|c|c|c|c|}
\hline
$T$~(MeV) & $\rho^{\rho+\omega}$~(fm$^{-3}$) & 
$\rho^{K^*}$~(fm$^{-3}$) & $\rho^{\phi}$~(fm$^{-3}$) &
$\langle \rho\sigma\rangle_T^{\rho+\omega}$~(fm$^{-1}$) &
$\langle \rho\sigma\rangle_T^{K^*}$~(fm$^{-1}$) & 
$\langle \rho\sigma\rangle_T^{\phi}$~(fm$^{-1}$) \\
\hline
$150$&$0.018$&$0.017$&$0.001$&$0.007$&$0.002$&$0.0007$\\
$165$&$0.037$&$0.035$&$0.003$&$0.016$&$0.006$&$0.002$\\
$180$&$0.068$&$0.064$&$0.007$&$0.035$&$0.014$&$0.005$\\
$195$&$0.115$&$0.108$&$0.012$&$0.070$&$0.030$&$0.010$\\
$210$&$0.184$&$0.172$&$0.021$&$0.131$&$0.059$&$0.020$\\
$225$&$0.279$&$0.260$&$0.034$&$0.231$&$0.108$&$0.039$\\
$240$&$0.404$&$0.377$&$0.052$&$0.388$&$0.188$&$0.070$\\
\hline
\hline
\end{tabular}
\caption{
Same as Table~I for vector particles.
} 
\end{center}
\end{table}
\begin{table}[htdp]
\begin{center}
\begin{tabular}{|c|c|c|c|}
\hline
$T$~(MeV) & $\epsilon^\pi$~(MeV/fm$^{-3}$) & $\epsilon^\eta$~(MeV/fm$^{-3}$) 
& $\epsilon^K$~(MeV/fm$^{-3}$)\\
\hline
$150$&$58.8$&$6.6$&$31.4$\\
$165$&$87.5$&$11.5$&$53.5$\\
$180$&$125.5$&$18.8$&$85.3$\\
$195$&$175$&$28.9$&$129$\\
$210$&$237$&$42.6$&$188$\\
$225$&$314$&$60.6$&$265$\\
$240$&$409$&$83.6$&$362$\\
\hline
\hline
\end{tabular}
\caption{
Energy density for the pseudoscalar particles in the gas.
} 
\end{center}
\end{table}
\begin{table}[htdp]
\begin{center}
\begin{tabular}{|c|c|c|c|}
\hline
$T$~(MeV) 
& $\epsilon^{\rho+\omega}$~(MeV/fm$^{-3}$) & 
$\epsilon^{K^*}$~(MeV/fm$^{-3}$)&$\epsilon^{\phi}$~(MeV/fm$^{-3}$) \\
\hline
$150$&$33.4$&$31.4$&$2.7$\\
$165$&$65.2$&$53.5$&$6.1$\\
$180$&$116$&$85.3$&$12.3$\\
$195$&$193$&$129$&$22.4$\\
$210$&$304$&$188$&$38.0$\\
$225$&$455$&$265$&$61.0$\\
$240$&$657$&$362$&$93.4$\\
\hline
\hline
\end{tabular}
\caption{
Energy density for the vector particles in the gas.
} 
\end{center}
\end{table}
\begin{table}[htdp]
\begin{center}
\begin{tabular}{|c|c|c|c|c|}
\hline
$T$~(MeV) & $\rho^{Tot}$~fm$^{-3}$ 
& $\langle \rho\sigma\rangle_T^{Tot}$~(fm$^{-1}$)
& $\epsilon^{Tot}$~(MeV/fm$^{-3}$) & g\\
\hline
$150$&$0.050$&$0.016$&$153$&$7.3$\\
$165$&$0.1$&$0.037$&$265$&$8.7$\\
$180$&$0.18$&$0.078$&$436$&$10.0$\\
$195$&$0.32$&$0.153$&$684$&$11.4$\\
$210$&$0.51$&$0.282$&$1030$&$12.9$\\
$225$&$0.77$&$0.492$&$1497$&$14.2$\\
$240$&$1.13$&$0.821$&$2112$&$15.5$\\
\hline
\hline
\end{tabular}
\caption{
Total number densities,  inverse absorption 
lengths, energy densities 
and number of active degrees of freedom, summed over 
the various components of the hadron gas.
} 
\end{center}
\end{table}

\section{Absorption in a Hagedorn gas}
In the range of temperatures we have found, one expects to deal with 
a hadron gas of increasing complexity, approaching the 
Hagedorn gas~\cite{hagedorn}, which has an infinite number 
of resonances with a level density 
exponentially increasing with mass. This situation gives rise 
to a limiting temperature~\cite{rafelski}, 
the Hagedorn temperature $T_{H}$, 
which was interpreted in~\cite{Cab&Par} as the temperature at 
which the transition to a 
quark-gluon plasma phase starts to take place. 
The Hagedorn temperature can be estimated from a 
fit to the resonance level density below $2$~GeV, appropriately 
smeared down in the lowest end. In \cite{rafelski} 
a value $T_{H}=158$~MeV is estimated. However, this value is 
lower than the observed hadron temperature at freeze-out and 
also lower than the estimates of the transition temperature 
from lattice QCD calculations, see Ref.~\cite{lattice}. As a reasonable 
compromise, we take $T_{H}=177$~MeV, which still fits the 
resonance spectrum and it agrees with lattice QCD calculations 
and with the observed freeze-out temperature. We use the 
partition function of the Hagedorn gas in the form~\cite{rafelski}:

\begin{equation}
\label{eq:partfunc}
\ln (Z_H)= V \left(\frac{T}{2 \pi}\right)^{3/2}\int_{0}^{\infty} 
\rho(m) e^{-m/T} dm =  V \left(\frac{T}{2\pi}\right)^{3/2}
\int_{0}^{\infty} C\frac{1}{(m_0^2+m^2)^{3/2}}
e^{m/T_H}e^{-m/T}dm,
\end{equation}
with~\cite{foot2}:

\begin{equation}
\label{eq:HGconst}
C=2.12~{\rm GeV}^2;~~~~~m_0=0.96~{\rm GeV};~~~~~T_H=177~{\rm MeV}
\end{equation}
and the energy density ($\beta=1/T$):
\begin{equation}
\label{eq:epsilonHag}
\epsilon=-\frac{\partial}{\partial \beta}\ln{\it Z}_{H}+\epsilon_\pi(T).
\end{equation}

To get the total energy density we have added the contribution 
of the pion, Eq.~(\ref{eq:enerd})
which is not included in $\rho (m)$. The number of 
effective degrees of freedom vs. 
temperature is shown in Fig.~\ref{f:gicombi}. 

\begin{figure}[ht]
\begin{center}
\epsfig{
height=7.truecm, width=8.truecm,
        figure=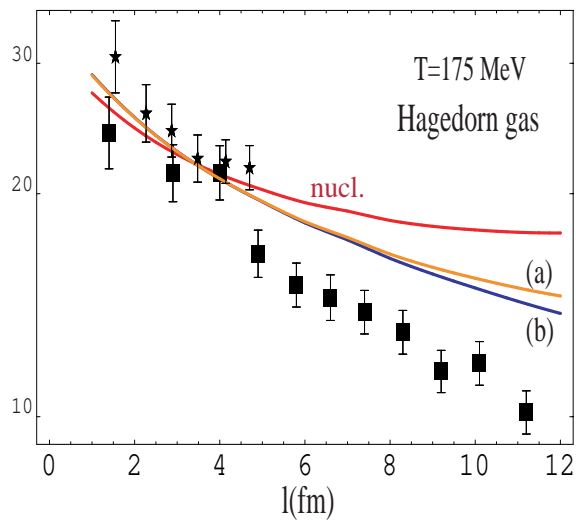
}
\caption{\label{f:hagedorn} \footnotesize 
The same as Fig.~\ref{f:abscombi}, without (a) or with (b) 
geometrical effects 
taken into account, using the energy-temperature dependence of 
the Hagedorn gas. The sharp rise of the degrees of freedom due 
to the vicinity of the Hagedorn temperature makes so that the 
temperature of the gas practically does not rise at all, the 
dissociation curve cannot become harder, and the prediction 
falls short from explaining the drop observed by NA50. 
Nuclear absorption alone is indicated by the upper curve.
}
\end{center}
\end{figure}

To make a numerical calculation possible, we 
have to make an assumption on the $\jj$ dissociation cross-section 
by the hadron resonances. We assume that 
only the pseudoscalar and the vector mesons we have considered 
before are relevant to dissociate the $\jj$. 
Starting from the results of Fig.~\ref{f:abscombi}, 
we extrapolate to 
increasing centrality using the energy-temperature relation 
of the Hagedorn gas. This is shown in Fig.~\ref{f:hagedorn}, 
using as initial temperature $T=175$~MeV. The result 
is quite spectacular. The sharp rise of the degrees 
of freedom due to the vicinity of the Hagedorn temperature 
makes so that the temperature of the gas practically does 
not rise at all, the dissociation curve cannot become harder, 
and the prediction falls really short from explaining the drop 
observed by NA50. The simplest interpretation of Fig.~\ref{f:hagedorn} 
is that with increasing centrality, more energy is deposited
but this goes into the excitation of more and more thermodinamical 
degrees of freedom leading to the final transition in the 
quark-gluon plasma. The curve shown in the figure would 
represent the limiting absorption from a hadron gas, 
anything harder being due to the dissociation of the $\jj$ 
in the quark-gluon plasma phase.   

Some words of caution are in order. In the framework of the CQM, 
it is certainly reasonable to expect the relevant insertions 
in the quark loop of Fig.~\ref{f:loop} to correspond to the Dirac matrices 
$S, P, A, V, T$ and the latter to be dominated by the lowest 
$q {\bar q}$, $S$-wave states we have been considering. On 
the other hand, we cannot exclude that decreasing couplings 
of the higher resonances may eventually resum up to a significant 
effect, which would change the picture obtained by truncating 
the cross section to the lowest levels. 

However, in all cases where this happens, like e.g. in deep 
inelastic lepton-hadron scattering, the final result reproduces 
what happens for free quarks and gluons. In our case, this would 
mean going over the Hagedorn temperature from the hadron into the 
quark and gluon gas, which is precisely what Fig.~\ref{f:hagedorn} 
seems to tell us. 

\section{Conclusions}

The prediction that its formation is 
suppressed in quark-gluon plasma~\cite{screening} 
makes the $\jj$ a very interesting probe for a possible 
phase transition in heavy ion collisions. When the 
prediction was made, it was believed that $\jj$ would 
suffer negligible absorption in the nuclear matter and 
in the hadron fireball formed during the collision.
Subsequent studies 
showed that this is not the case and, for the probe to be efficient,
one must determine in a reliable way the absorption of $\jj$ 
from these two ``conventional'' sources. The absorption length 
of $\jj$ in nuclear matter has been determined~\cite{nuclabs} 
from the the cross-section of $p+A \to \jj +$anything as a 
function of the nucleon number, $A$. 

A method to estimate the absorption from the comoving particles 
was presented in a previous paper~\cite{ioni}, under the simplified 
assumption that the comoving particles are a thermalised pion gas 
at temperature $T$. Starting from a calculation of the $\pi + 
\jj \to D^{(*)}\bar D^{(*)}$ cross-section based on the Constituent 
Quark Model~\cite{cqm}, we showed that the absorption is potentially 
large and strongly temperature dependent. $T$ itself can be determined 
by a fit to the data at low centrality, where the pion gas 
approximation could be better justified. Comparing the extrapolation 
to larger centralities with data, one could then judge if a further 
suppression is at work. The temperatures found in~\cite{ioni}, 
$T \approx 200$~MeV, were rather on the high side but still in 
the right ball-park, a reassuring indication for a calculation 
completely based on microscopic parameters. On the other hand, 
for the method to be reliable, it is necessary to apply a better 
approximation than the simple pion gas.

In the present paper we have  extended the calculation to the 
lowest-lying pseudoscalar and vector mesons, the $S-$wave 
$q {\bar q}$ states. Contributions from higher resonances like 
the $A_1$ and the other $P-$wave $q {\bar q}$ states could also 
be considered, at the price of increasing theoretical uncertainties, 
and should not change the picture. Vector mesons give substantial 
contributions to $\jj$ dissociation, due to the higher multiplicities 
and because they are close or above threshold. As a consequence, 
we found a larger absorption in the heath bath at temperature $T$ 
and a much lower range of temperatures to fit low 
centrality data, $T=165-185$~MeV. This range agrees with the 
temperatures found from (i) particle abundances in ion collisions 
at freeze-out, (ii) the transition temperature found in lattice 
QCD calculations and (iii) the limiting temperature of a hadron 
gas~\cite{hagedorn, rafelski}, as estimated from the experimental 
hadron level density.

In the range of temperatures considered, the number of active 
degrees of freedom is rapidly increasing, due to the excitation 
of more and more hadron levels (Fig.~\ref{f:gicombi}). 
This concept provides the ground 
for the existence of a limiting temperature in a hadron gas, the 
Hagedorn temperature. For the gas made by the comoving 
particles, the increase of the degrees of freedom ``blocks'' the 
temperature to the values observed for the low centrality events 
and prevents the opacity of the fireball to $\jj$ to increase further. 
In fact, assuming a limiting temperature of about $170-180$~MeV 
from the outset, the decrease of $\jj$ production observed by NA50 
at large centralities cannot be explained with absorption by either 
the nuclear matter or the comoving particles, as shown in 
Fig.~\ref{f:hagedorn}.  

We cannot exclude that an infinite tower of higher resonances 
with decreasing couplings 
could eventually change the 
picture obtained by truncating the cross section to the lowest 
levels. Should this happen, however, we would expect the result 
to be close to that obtained by replacing the resonances with 
quarks and gluons, i.e. the situation one encounters in the QGP.

In conclusion, we believe that our calculation has produced 
a reliable estimate of $\jj$ absorption in a hadron gas where 
only the lowest resonances are excited. It can agree marginally 
with the data at large centrality, at the expenses of accepting 
temperatures which are quite high ($T \approx 200$~MeV) with 
respect to the current theoretical estimates of the limiting 
hadron temperatures. On the other hand, should one introduce 
the limiting temperature as a boundary condition, it is 
impossible to fit the data at large centrality, thus providing 
considerable support to the idea that the suppression is produced 
by the quark-gluon plasma.

Further progress can be envisaged in several directions. Further 
measurements at low energy, aimed at resolving the experimental issue of 
the relative normalization of S-U vs. Pb-Pb 
data, would restrict the range of temperatures required at 
low centrality and allow for a more precise extrapolation 
with respect to what shown in Fig.~\ref{f:abscombi}. The 
onset of other possible signatures of the phase transition, 
such as strangeness production, should be correlated to the present 
signal, in centrality and energy. Finally, a quantitative calculation 
of the expected absorption curve, should the QGP be formed, 
would be of crucial importance.
 
\section*{Acknowledgements}

We have profited from discussions with several colleagues. 
In particular, we would like to thank F.~Antinori, G.~Bruno, 
R.~Fini, B.~Ghidini, 
C.~Louren\c{c}o, B.~M\"{u}ller, H.~Specht, U.~Wiedemann for interesting 
discussions and P.~Giubellino, L.~Ramello and E.~Scomparin for 
intruducing us to the most recent NA50 data. 
FP thanks the CERN-TH Unit for partial financial support. 
ADP thanks the Physics Department of the Bari University for 
kind hospitality. 
The kind hospitality by the CERN-TH Unit and the exciting
environment of the Heavy Ion Forum are gratefully acknowledged.


\begin{thebibliography}{99}
\bibitem{nota2d} This range of temperatures is subject to a very tiny 
variation when the geometrical factor in Eq.~(\ref{eq:abscomov}) is taken
to be $8/3\pi R$ instead of $3/4 R$, i.e., if we consider a bidimensional 
(circle) fireball of radius $R$ instead of a spherical one.
\end{thebibliography}
\end{document}